\documentclass[12pt]{article}
\usepackage[margin=1in]{geometry}
\usepackage{graphicx}
\usepackage{rotate}
\usepackage[sort&compress,numbers]{natbib}

\newcommand{\panda}{\ensuremath{_{PAnDA}}}

\begin{document}

\title{On the feasibility and utility of exploiting real time database
  search to improve adaptive peak selection}
\author{Benjamin Diament$^1$, Michael J.\ MacCoss$^2$ and William
  Stafford Noble$^{2,1}$\footnote{To whom correspondence should be
    addressed: william-noble@uw.edu}}
\date{}
\maketitle

{\bf Running head:} Real time database search to improve adaptive peak
selection

\footnotetext[1]{Department of Computer Science and Engineering, University of Washington}
\footnotetext[2]{Department of Genome Sciences, University of Washington}

\begin{abstract}

{\bf Rationale}: In a shotgun proteomics experiment with
data-dependent acquisition, real-time analysis of a precursor scan
results in selection of a handful of peaks for subsequent isolation,
fragmentation and secondary scanning.  This peak selection protocol
typically focuses on the most abundant peaks in the precursor scan,
while attempting to avoid re-sampling the same m/z values in rapid
succession.  The protocol does not, however, incorporate analysis of
previous fragmentation scans into the peak selection procedure.

{\bf Methods:} In this work, we investigate the feasibility and
utility of incorporating analysis of previous fragmentation scans into
the peak selection protocol.  We demonstrate that real-time
identification of fragmentation spectra is feasible in principle, and
we investigate, via simulations, several strategies to make use of the
resulting peptide identifications during peak selection.

{\bf Results:} Our simulations fail to provide evidence that peptide
identifications can provide a large improvement in the total number of
peptides identified by a shotgun proteomics experiment.

{\bf Conclusions:} These results are significant because they point
out the feasibility of using peptide identifications during peak
selection, and because our experiments may provide a starting point
for others working in this direction.
\end{abstract}

\clearpage

\section*{Introduction}

In broad terms, the goal of many tandem mass spectrometry experiments
is to identify, from a given complex mixture, as many distinct
peptides as possible in an efficient fashion.  Accordingly, many
aspects of a typical shotgun proteomics workflow are tuned to optimize
the number of high quality spectra derived from distinct peptides.
For example, the liquid chromatography phase provides the mass
spectometer, at each time point, with a sample of greatly reduced
complexity relative to the initial sample, thereby reducing the probability of observing
hard-to-identify chimeric spectra, derived from a population of two or
more peptide species.  Similarly, the flow of sample off the liquid
chromatography column is calibrated so that each MS1 scan over intact
peptides is followed by sufficient MS2 scans to characterize a large
proportion of the identifiable peptide species.

In this work, we focus on a particular component of this workflow,
namely, the selection of peaks within an observed MS1 spectrum for
subsequent isolation and characterization via MS2 scans.  With each
MS1 scan, the spectrometer's on-board computer must select a subset of
peaks to isolate, fragment, and measure in an MS2 scan. This task is
challenging because, in each duty cycle, hundreds of MS1 peaks may be
reasonable candidates for MS2 analysis, but the available time
typically permits only a small number (5--10) to be analyzed.
Ideally, a peak within an MS1 spectrum indicates the presence of a
homogeneous population of peptides, and the peak height can be used as
a proxy for peptide abundance.  Existing protocols therefore use this MS1
information to select a set of $m/z$ windows for subsequent
fragmentation and MS2 characterization.

Today, the most common peak selection strategy is based upon the
notion of an {\em exclusion list}, which is a dynamically maintained
list of m/z values that are excluded from being selected for
fragmentation \cite{kohli:alternative}.  The standard exclusion list protocol
works as follows. A list of maximum fixed size $m/z$
values (in the range of
50--500, depending on the instrument) is maintained in the on-board computer's memory and is
initialized to be empty. With each MS1 scan, the highest-intensity
peaks are examined in order. Any peak already on the exclusion list is
ignored. Any peak not on the exclusion list is selected for
fragmentation and is placed on the exclusion list. This process
continues until the start of the next duty cycle.  Once a peak is
placed on the exclusion list, it remains there for a user-specifiable
period of time, typically around 30 seconds.

The main goal of the exclusion list protocol is to maximize the number
of distinct peptides identified, which it does using a two-part
approach. First, the protocol aims to gather spectra principally from
abundant peptides, which are likely to be of higher quality and hence
easier to identify.  Second, the exclusion list attempts to avoid
re-sampling the same peptide species twice, thereby ensuring the
selection of a diverse set of peptide species. Taken together, these
goals are intended to lead to a broad sampling across abundant
peptides in the sample.

A variety of methods have been proposed for improving upon the
standard exclusion list by performing analyses interleaved between
repeated rounds of MS experiments.  Thus, peptides identified in an
initial round are excluded from analysis in subsequent rounds
\cite{chen:enhanced, wang:exploring, bendall:enhanced,
  scherl:nonredundant} or precursor peaks identified in an MS analysis
are used to select peaks for analysis in a subsequent MS/MS experiment
\cite{rinner:integrated, picotti:implications, schmidt:integrated,
  zerck:iterative, hoopmann:post}.  Both types of approach yield a
larger number of total identifications overall.

In contrast, we consider the possibility of augmenting the peak
selection protocol with information about whether a selected MS1 peak
was successfully identified {\em in the current experiment}.  Such an
approach is feasible because we have recently demonstrated that
spectrum identification via database search can be carried out
extremely rapidly \cite{diament:faster}---at a rate of 1550 spectra/s
for a fully tryptic digestion of proteins from a complex eukaryote.
This high-speed search procedure raises the possibility of performing
real-time spectrum identification and exploiting the resulting
identifications in the context of a peak selection protocol.

In this work, we first establish the technical feasibility of
real-time spectrum identification.  We then explore the hypothesis
that the information provided by real-time identification can be
successfully exploited by an adaptive peak selection protocol to yield
better peptide identifications over the course of a shotgun proteomics
experiment.  We refer to such a protocol as {\em ID-informed adaptive
  peak selection}.  However, prior to implementing real-time
identification software and actually coupling it with an on-board peak
selection protocol, we first chose to simulate a variety of
ID-informed adaptive peak selection schemes.  Unfortunately, these
simulations did not strongly support our initial hypothesis: among
many algorithms that we tried, only a few led to an increase in the
estimated number of distinct peptides identified, and the increases
were generally quite modest.

This article can therefore be read in several ways: as a statement of
work in progress, as a cautionary tale, or as a stimulant for other
researchers working in this area.  On the one hand, we are quite
certain that real-time spectrum identification is technically
feasible.  On the other hand, we cannot at this time provide strong
evidence for the utility of such an approach.

\section*{Methods}

\paragraph{Data}
In our simulation, we pooled MS2 data from a previously described
collection of 11 replicate experiments \cite{hoopmann:post}. Briefly,
{\em C.\ elegans} were grown on enriched peptone plates seeded with
the OP50 strain of E. coli at 20°C. After lysing and centrifugation,
the lysate was denatured using 0.1\% RapiGest SF (Waters Corporation,
Milford, MA) in 50 mM ammonium bicarbonate pH 7.8.  The resulting {\em
  C.\ elegans} digest (4 µg) was loaded from the autosampler onto a
75-$\mu$m capillary column placed in line with a Waters NanoAcquity
HPLC and autosampler.  Peptide elution was performed using two buffer
solutions, as described previously. Tandem mass spectra were acquired
using either traditional data-dependent acquisition with dynamic
exclusion turned on or PAnDA.  In PAnDA, MS/MS scans are performed on
ions selected from an m/z inclusion-exclusion list computed to prefer
ions, which, based on prior analysis of the $\mu$LC-MS on the same
sample, were expected to be abundant peptides that had not been
selected in earlier replicate experiments.  In both cases, a single
high resolution mass spectrum was acquired at 60,000 resolution (at
m/z 400) in the Orbitrap mass analyzer in parallel with 5 low
resolution MS/MS spectra acquired in the LTQ.

\begin{table}
\centering
\small
\begin{tabular}{lrr}
Experiment & Spectra & IDs \\
\hline
e10 & $7243$ & $2007$ \\
e10\panda & $6807$ & $1370$ \\
e11 & $7226$ & $1863$ \\
e11\panda & $6637$ & $1048$ \\
e12 & $7136$ & $1650$ \\
e12\panda & $6496$ & $845$ \\
e14 & $6666$ & $1478$ \\
e14\panda & $6275$ & $1017$ \\   
e15 & $6765$ & $1290$ \\         
e15\panda & $6476$ & $1109$ \\   
e16 & $7054$ & $1656$ \\         
\end{tabular}
\caption{{\bf Data sets}  The table lists, for each of the 11 experiments, the number of MS2 spectra and the number of spectra that were successfully identified by Tide+q-ranker using a $q$-value threshold of $0.01$.
\label{table:data}}
\end{table}

\paragraph{Searching MS2 data}
Each experiment was searched using Tide \cite{diament:faster} against
a forward and a reversed protein database consisting of the predicted
open reading frames from {\it C.\ elegans} and common contaminants
(Wormpep v160, 27,499 proteins). Tide was run with the following
parameters: (1) fully-tryptic digestion of database proteins; (2)
peptide lengths between $6$ and $50$ residues; (3) peptide masses
between $200$ and $7200$ Da; (4) static cysteine modification of
$+57.02146$ Da. (carboxamidomethylation); (5) mass tolerance of $\pm
3.0$ Da.; (6) monoisotopic precursor.  The Tide results were further
processed using q-ranker \cite{spivak:improvements} with the default
parameters.  Peptide-spectrum matches were accepted as ``correct''
using a $q$-value threshold of $0.01$.  For the simulation results
shown in Figure~\ref{figure:baseline}--\ref{figure:simulation}, a
peptide was considered to be identified if at least one PSM mapped to
it.  Table~\ref{table:data} provides a summary of each data set,
including the number of spectra and the number of PSMs at $q\leq
0.01$.

\paragraph{Aligning MS1 data}
To pool the MS2 data required first aligning the MS1 data from
experiment e10 with respect to each of the 10 experiments.  This
alignment was done by considering, for a given pair of experiments, all
peptides that were succesfully identified (using a $q$-value threshold
of $0.1$) exactly once in each of the two experiments. The
pairs of retention times were then used in a regression minimizing
perpendicular least squares.  The resulting
regression line allowed mapping an observed retention time in a given
experiment to its corresponding retention time in experiment e10.

\section*{Results}

\subsection*{Requirements for real-time peptide identification}
\label{section:lowlatency}

We begin by establishing the technical feasibility of real-time
spectrum identification.  Before doing so, however, we introduce a few
concepts that are necessary to understand the discussion.  Any
database search solution to spectrum identification is based on the
idea of enumerating, for each observed spectrum, all of the {\em
  candidate peptides} whose intact mass is close to the inferred mass
associated with the spectrum.  Here, ``close to'' is measured in
Daltons, and is a parameter set by the user.  In the SEQUEST
algorithm, which we focus on below, each candidate peptide generates a
corresponding theoretical spectrum.  A score function, called XCorr,
computes the quality of the match between an observed spectrum and a
theoretical spectrum.  Note that, although the original SEQUEST
algorithm used a two-pass scoring scheme involving a prelimimary score
(Sp) and XCorr, more recent implementations omit the Sp scoring
\cite{eng:fast}.  SEQUEST thus returns, for each spectrum, the
candidate peptide whose theoretical spectrum matched with the highest
XCorr value.

To achieve real-time peptide identification, we need a system that can
do three things, of which a typical peptide database search program
can do only the first: (1) Perform a database search of reasonable
complexity, (2) achieve high bandwidth on a single computer, and (3)
achieve low latency between spectrum acquisition and identification.

To achieve reasonable search complexity for a typical shotgun
proteomics experiment, we want to be able to search a database of
proteins from a single target organism, plus contaminants, with some
reasonable settings, such as a semi-tryptic digest and a sufficiently
wide precursor tolerance window. Reasonable database size, digest, and
tolerance window are significant contributors to search time. All
modern peptide search software, can perform reasonably complex
searches.

\begin{table}
\small
\centering
\begin{tabular}{lrrrr}
& \multicolumn{2}{c}{Semi-tryptic}
& \multicolumn{2}{c}{Fully tryptic} \\
& $\pm 3.0$ & $\pm 0.25$ & $\pm 3.0$ & $\pm 0.25$ \\
\hline
{\em S.\ cerevisiae}  & 536  & 1198 & 2738 & 3244 \\
{\em C.\ elegans}     & 77.2 &  413 & 1554 & 1980 \\
\end{tabular}
\caption{{\bf Throughput of the Tide database search software} Each
  entry in the table represents the number of spectra/s analyzed by
  Tide using the specified database and search parameters.
  \label{table:tide}}
\end{table}

For the second item in the list above, we have recently described Tide
\cite{diament:faster}, which is a reimplementation of the SEQUEST
algorithm \cite{eng:approach} that achieves extremely high bandwidth
on a single computer.  Table~\ref{table:tide} shows Tide's throughput,
measured in spectra per second on a single thread of a single CPU, for
eight different experimental settings: using a protein database
derived from {\em S.\ cerevisiae} or {\em C.\ elegans}, generating
peptides using a fully tryptic or semi-tryptic digestion, and
selecting candidate peptides from the database using a large (3.0 Da)
or small (0.25 Da) precursor window.  Even for the most demanding
setting ({\em C.\ elegans}, semi-tryptic, 3.0 Da), Tide's throughput
is approximately an order of magnitude higher than the 5--10 spectra/s
that would be required for a real-time pipeline.

However, the current implementation of Tide does not achieve the third
aim listed above.  Instead, Tide assumes that all spectra to be
searched are available at the start of computation. This assumption,
while reasonable in the context of most proteomics laboratory
pipelines, fails in the context of real-time peptide identification.

To achieve ID-informed adaptive peak selection, we need to identify
spectra in time to make on-the-fly decisions on which peaks to select
for fragmentation.  We therefore need a low-latency peptide
identification method, meaning that spectra are identified within a
very short time (say, a few seconds) after the spectrum has been
acquired. This requirement is distinct from that of high bandwidth,
though the two requirements are related and can be addressed using
many of the same approaches.

\subsection*{Adapting Tide for real-time computation}

Tide begins processing by sorting the input spectra by neutral mass. A
low-latency version of Tide would not be able to access all spectra at
the start of processing, and hence cannot this initial sort. Some
alternative is required, and we discuss this next.

The initial sort of spectra, coupled with a pre-sorted database of
peptides, achieves two important objectives: avoiding redundant
computation by computing a theoretical spectrum for the same peptide
in the context of identifying two different spectra and maintaining a
small memory footprint over the course of computation. The first
objective is critical for speed, and some alternative implementation
is required to achieve the same level of performance.  However, the
second objective---maintaining a small memory footprint---is of
intermediate importance: there needs to be sufficient memory available
to complete the computation, but there is no need to conserve beyond
that point.

A simple substitute strategy for the initial sort is to preload all
theoretical spectra in memory in advance of collecting the first
spectrum. Before any spectrum is processed, the index of peptides is
loaded from disk and the accompanying data structures are all stored
in memory. This preloading approach has the potential for enormous
impact on the program's memory usage, but it also leaves the total
running time (including the preloading step) almost exactly the same
as for the current Tide. This approach maintains the guarantee that
each theoretical spectrum will be computed at most once, at program
startup. Thereafter, the theoretical spectra will not have to be
recomputed at all. All the functions of Tide are performed under the
preloading approach, but in a different order than for Tide. Except
for the larger memory footprint, the resource usage does not change.

We now address the question of whether enough memory would be
available to perform a reasonable peptide identification experiment
under the preloading strategy. We first measure empirically how much
memory would be consumed. The current implementation of Tide
interleaves the preprocessing of the observed spectra, the computation
of the theoretical spectra, and computing the XCorr. The interleaving
allows us to free memory used by the theoretical spectrum computation
once the relevant candidate peptides are no longer needed. If we do
not free the corresponding memory, then we can measure how much memory
would be used by the alternative approach described.

\begin{figure}
\centering
\includegraphics[width=4.0in]{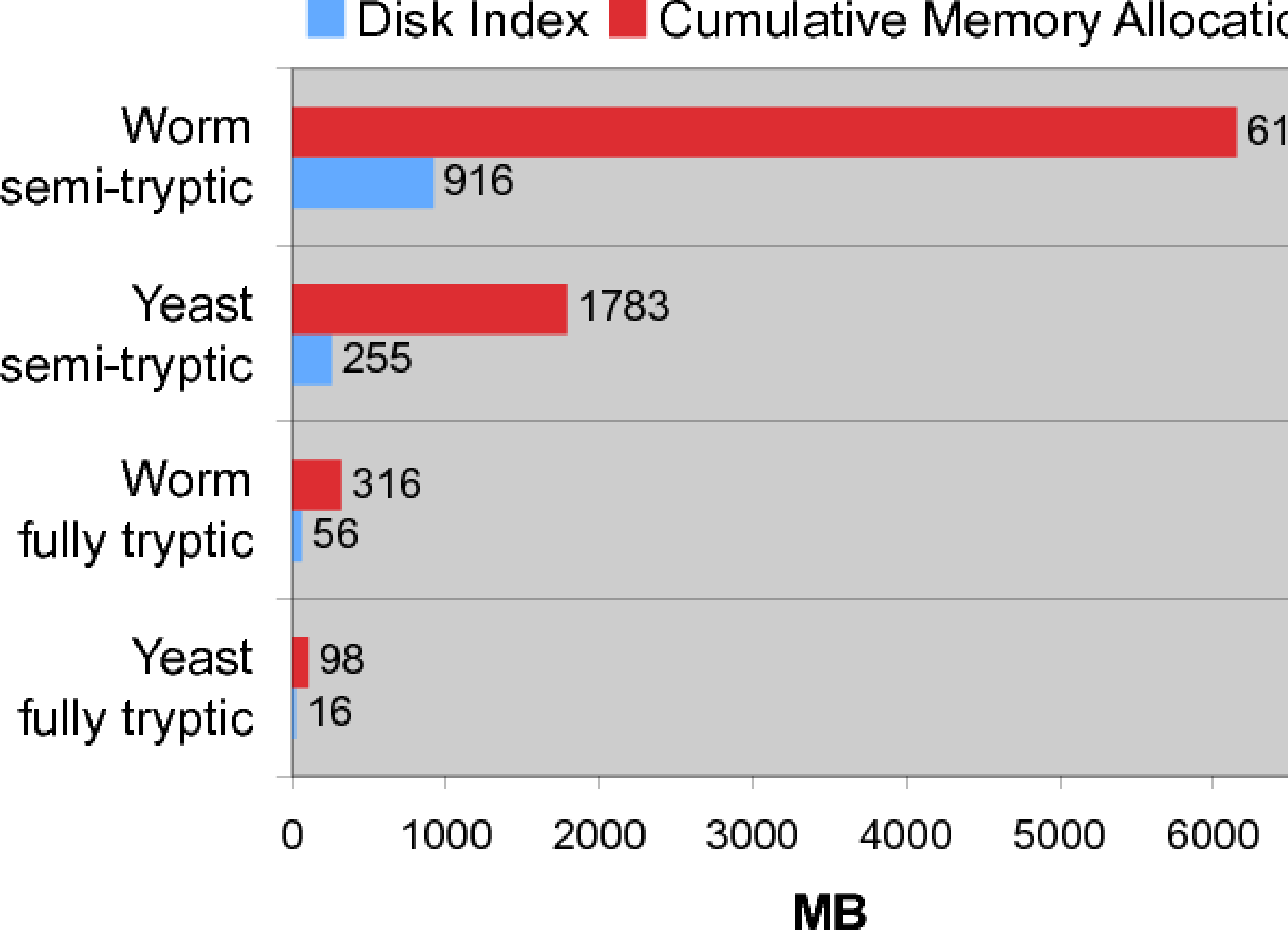}
\caption{{\bf Disk and memory footprint for preloaded theoretical
    spectra.} For each of four benchmark data sets, the size of the
  index on disk is shown. We also show the cumulative amount of memory
  allocated during processing (ignoring deallocations) for each
  benchmark. The resulting value is an upper bound on the memory required to
  preload the machine with all theoretical spectra.
  \label{figure:footprints}}
\end{figure}

Figure~\ref{figure:footprints} shows the total disk usage and
cumulative memory allocated (ignoring reuse) by Tide when run in each
of four modes. We see that the worm dataset, which needs the larger
database, requires more memory than yeast under comparable settings.
We also see that semi-tryptic digestion is considerably more
memory-intensive than fully-enzymatic digestion. However, even in the
most intensive case, only 6GB of memory is allocated in total. These
results show that on a modern workstation, there is sufficient memory
not only for a fully tryptic search, but even for a substantially
larger semi-tryptic search, even without substantial changes to data
structures.

We have not implemented a low-latency version of Tide according to the
method described here.  We opted first to perform simulations to
ascertain the feasibility of ID-informed adaptive peak selection. Had
we seen more positive results from testing, then a low-latency
implementation of Tide would have been the next step.

\subsection*{Simulation protocol}

Before attempting to implement ID-informed adaptive peak selection in
the context of a real shotgun proteomics experiment, we wanted to
prove the concept first. To this end, we simulated the action of a
spectrometer able to select peaks adaptively. In the simulation, the
spectrometer has access to real-time peptide identifications as they
would be performed by a real-time Tide implementation and is able
accordingly to select peaks for fragmentation.

\begin{figure*}
\centering
\begin{tabular}{cc}
\includegraphics[width=3.0in]{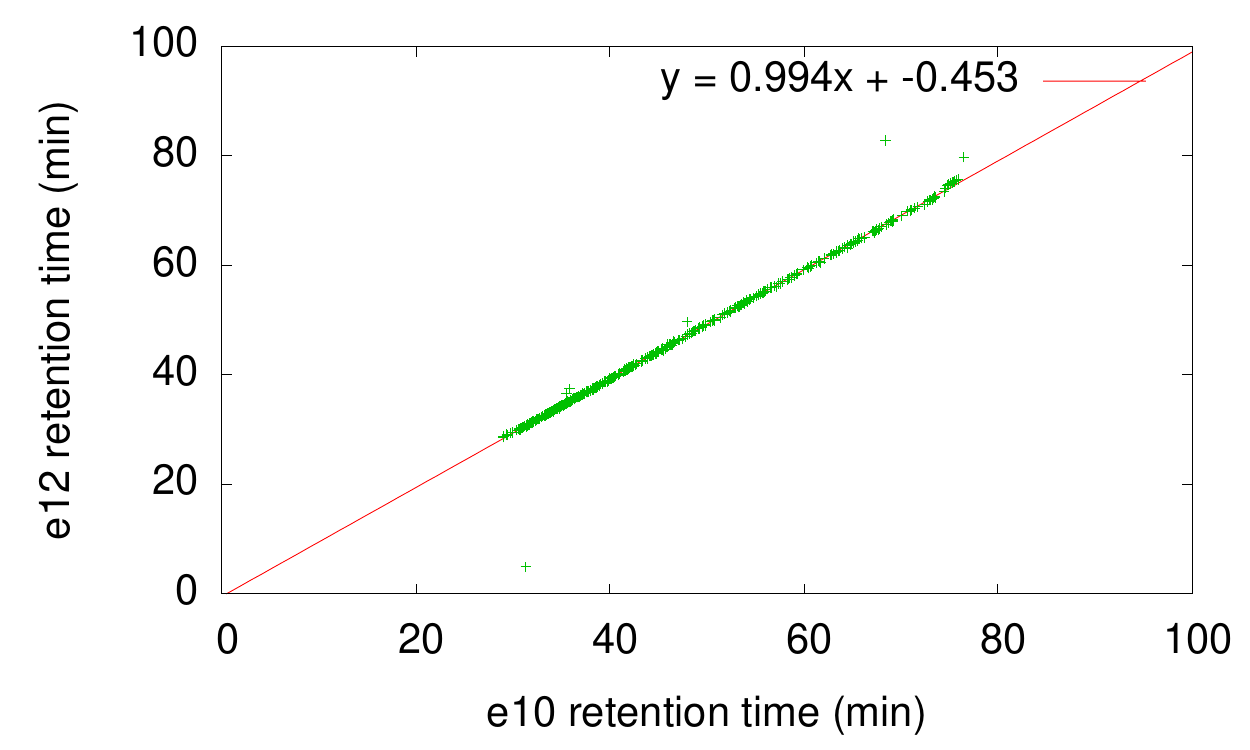} &
\includegraphics[width=3.0in]{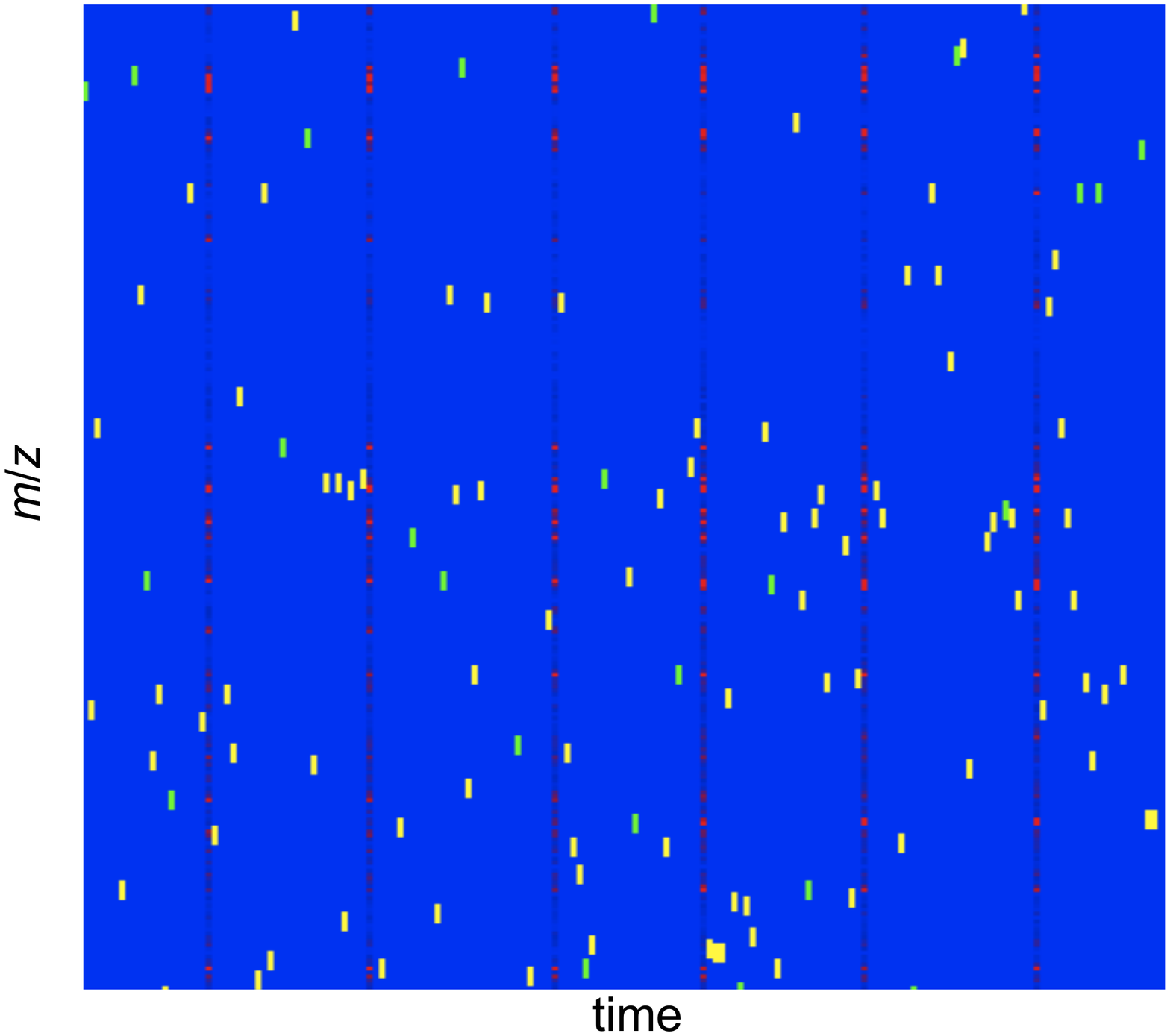} \\
\end{tabular}
\caption{{\bf Simulation data}.  (A) Retention-time alignment of
  replicate experiments e10 and e12. Each green point in the figure
  corresponds to a peptide identified in both experiments at $q<0.1$.
  The diagonal line is the result of a regression to the observed
  points (see Methods for details). (B) MS1 data interleaved with
  combined MS2 data from multiple experiments. A tiny region of the
  total data is shown magnified (about 7 seconds and 2 $m/z$ from
  experiment e10, which ran 100 minutes and covered 1000 $m/z$). The
  vertical red lines represent the MS1 data from experiment e10, with
  higher-intensity peaks indicated in brighter shades of red. A green
  tick mark shows peaks picked for analysis during the actual run of
  e10. A yellow tick mark shows where interpolated MS2 data from
  another replicate experiment is available during simulation.  the
  experiments.
  \label{figure:simdata}}
\end{figure*}

In a perfect simulation, an MS2 scan would be available for any MS1
peak that a peak selection algorithm might choose. To approximate this
setting in our simulation, we pooled MS2 data from a previously
described collection of 11 repicate experiments \cite{hoopmann:post}
(see Methods).  Pooling the data required first aligning the
replicate experiments by retention time. One experiment, called {\it
  e10}, was used as the basis for MS1 data that the simulator would
give to the candidate peak picker. Then the MS2 data was taken from each
of the other 10 experiments and compared with the MS2 data of
e10. Figure~\ref{figure:simdata}(A) shows the computed retention-time
alignment between e10 and another replicate, e12, as an example. Such
an alignment allows us to map any identified MS1 peak onto the
baseline experiment e10.  Figure~\ref{figure:simdata}(B) shows the
result graphically: MS1 data from experiment e10 is interleaved with
MS/MS data from all replicates. The hope is that if a candidate peak
picker requests an MS/MS scan of a particular peak and it wasn't
selected in the actual run of e10, then one of the other experiments
may have taken a scan of the same peak at about the same time.

\begin{figure}
\centering
\includegraphics[width=4.0in]{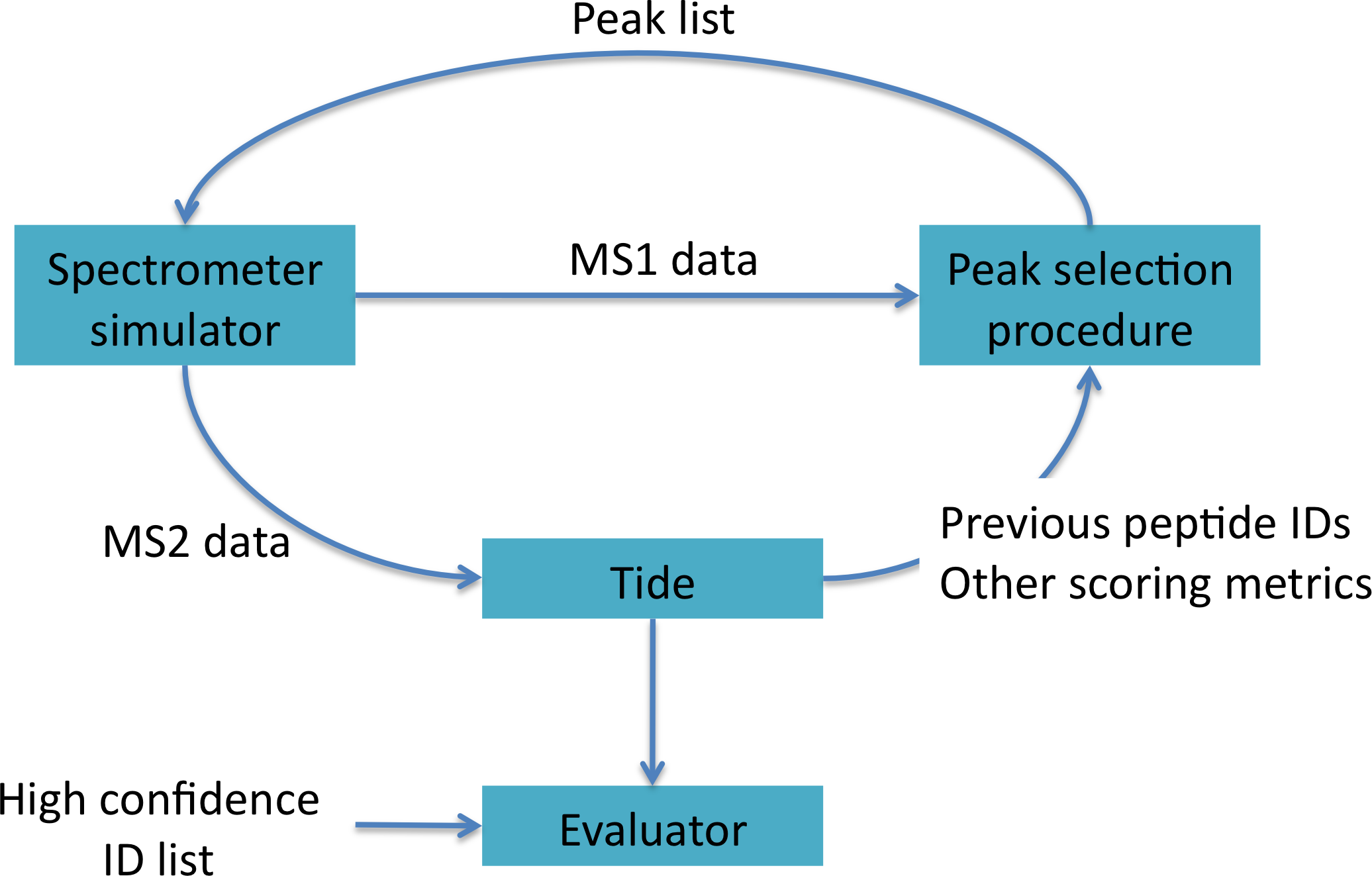}
\caption{{\bf Structure of spectrometer
    simulator} The spectrometer simulator component runs through all the MS1
  scans from experiment e10, one at a time. It passes this MS1 data to a modular
  peak selection procedure, which selects peaks for fragmentation. The simulator
  checks whether it has a corresponding MS/MS scan. If it does, then it uses
  Tide to retrieve the corresponding identification and XCorr values. The
  information from Tide is also passed to the peak selection procedure. Finally,
  the list of identified peptides is sent to the evaluator for comparison
  against a precomputed list of high-confidence identifications.
  \label{figure:simulator}}
\end{figure}

Given this data, the simulator works in conjunction with a given
experimental peak-selection procedure, Tide and an evaluator module,
as shown in Figure~\ref{figure:simulator}. We implemented in Python a
framework for simulation according to the diagram. The spectrometer
simulator, the Tide results module, and the evaluator were implemented
in one Python program which was allowed to run with any
separately-supplied peak picker procedure. This approach enabled us to
run simulations under varying parameter settings.

After completion of one simulation run, we are primarily interested in
the total number of unique peptides identified over the course of the
entire spectrometry experiment. However, because of the limits
inherent in simulation---in particular, the simulator's inability to
return every MS2 spectrum requested by the peak picker---we could not use this simple
metric to make reasonable comparisons against the basic exclusion
list, for which complete corresponding MS2 data is available.  We
therefore selected a two-dimensional metric: the number of unique
peptides identified versus the number of spectrum requests for which
an MS2 scan was available.

\subsection*{Performance of the baseline method}

\begin{figure}
\centering
\includegraphics[width=4.0in]{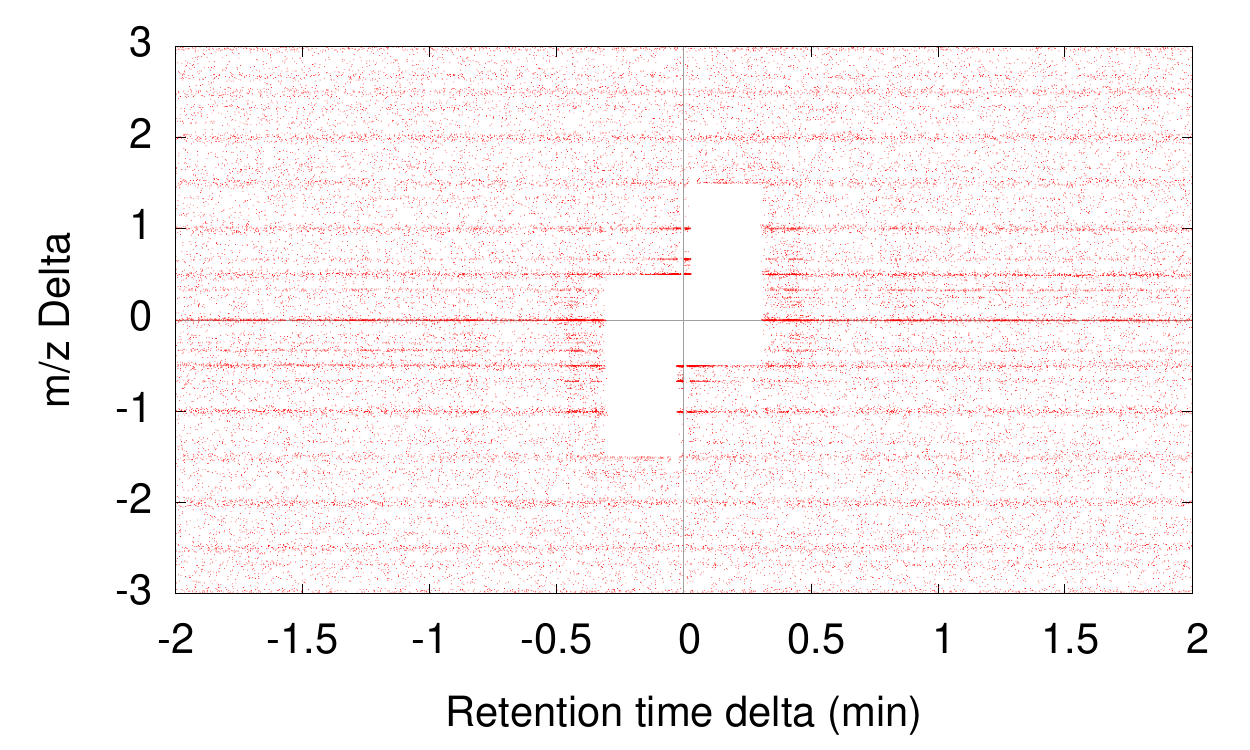}
\caption{{\bf Pairwise differences between peaks in experiment e10.}
  The figure plots the differences in retention time and $m/z$ between
  all pairs of nearby peaks that were actually selected for
  fragmentation in experiment e10. Thus, the figure illustrates the
  effect of the on-board exclusion list.
  \label{figure:deltas}}
\end{figure}

To establish a baseline for our simulation, we began by attempting to replicate, as closely as possible, the behavior of the actual exclusion list that was used to generate our experimental data.  The basic settings included an exclusion duration of  $18$ seconds and an exclusion window of $\Delta m/z \in [-0.5, 1.5]$ relative to any excluded peak.  That is, when a candidate peak is considered for fragmentation, it is checked
for two criteria against every other peak previously chosen: if the candidate
occurs within 18 seconds of a previously chosen peak and also falls within the
exclusion window of that peak by $m/z$, then the candidate may not be picked. A
picture of these settings can be drawn by plotting the differences between all
pairs of selected peaks within e10, by precursor $m/z$ and retention time
(Figure~\ref{figure:deltas}). Such a picture is symmetric by construction, and
the blank region highlights the excluded differences. This picture shows that
our settings for the exclusion window and exclusion duration are correct.

\begin{figure}
\centering
\includegraphics[width=3.0in]{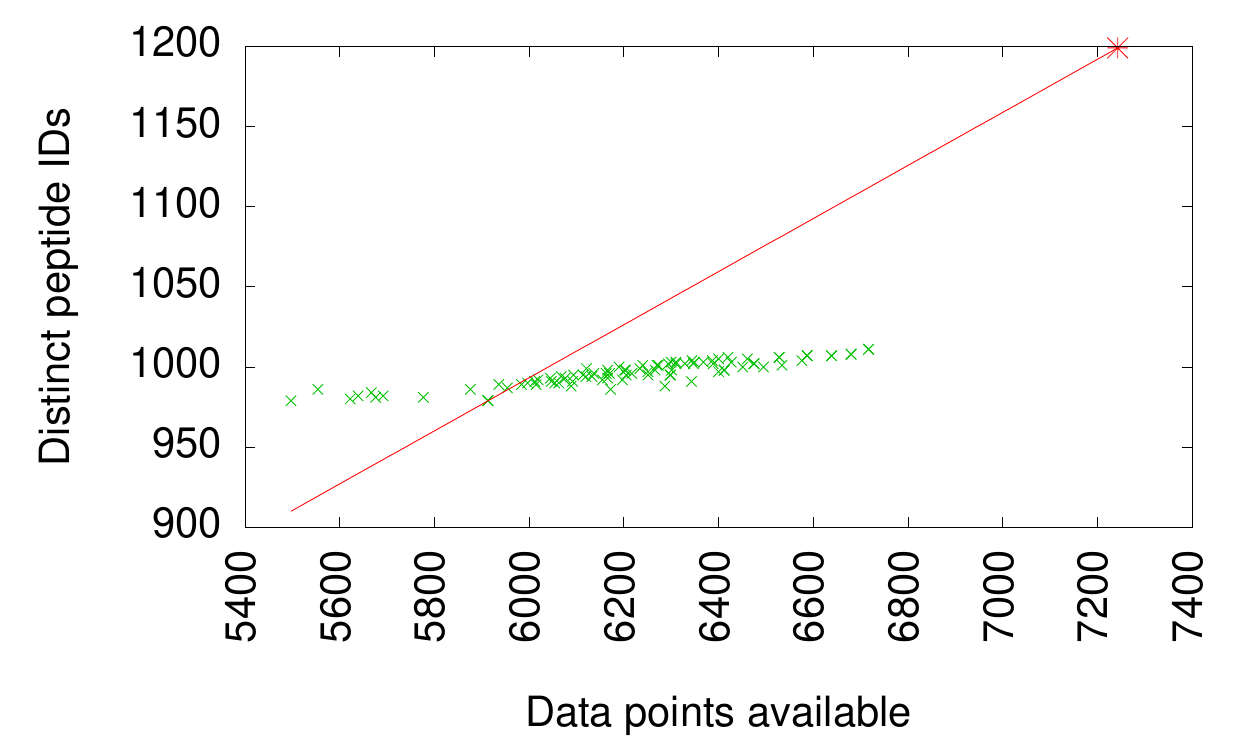}
\caption{{\bf Baseline performance of the simulated exclusion list.}
  In the figure, each point represents a complete simulated mass
  spectrometry experiment under varying parameterizations of the
  simulator. The figure plots the total number of distinct peptides
  identified as a function of the number of selected peaks for which
  the simulation successfully returned an MS2 spectrum. The point in
  the upper right represents the actual e10 experiment, and the line
  represents the ratio of identifications to total peaks acquired
  throughout the run of e10.
  \label{figure:baseline}}
\end{figure}

The results of this initial simulation are shown in
Figure~\ref{figure:baseline}.  In the figure, each point corresponds
to a different setting of simulation parameters, where the parameters
define the tolerance of the simulator for returning MS2 data whose
retention time or $m/z$ precursor differed from that requested.  The
most striking observation to be drawn from
Figure~\ref{figure:baseline} is that the simulated exclusion list
yields far fewer MS2 spectra and, correspondingly, fewer distinct
peptide identifications than the real experiment (shown as a point in
the upper right corner of the plot).  Even if we select the simulation
parameters that yield the best results ($m/z$ tolerance of $0.4$
Da/charge and retention time tolerance of $1.8$ minutes), the
simulator only successfully returns $6717$ MS2 spectra, which is $526$
fewer spectra than were actually produced in experiment e10.  On the
other hand, the rate of successful identifications is sometimes higher
and sometimes lower than in the real experiment, as shown by the
relative locations of the green simulation points and the red diagonal
line.

We hypothesize that the tendency to return fewer MS2 spectra in the simulation arises for two reasons.  First, we observed that, in some cases, the on-board computer selected peaks that are not taken from the top few peaks
by intensity. Although some of the skipped-over peaks could be explained by their presence on the exclusion list, not all of the skipped peaks could be ``explained away'' in this fashion.  We therefore suspect that the
actual peak selection algorithm includes some features of which
we were not aware. Second, we observed that the exclusion list is surprisingly
unstable, in that slight perturbations in the choice of peaks early in the
experiment lead to far-reaching differences in the choice of peaks later
on. This phenomenon appears to be the result of a feedback effect between the
peak selector and the exclusion list over the course of a run: a change in peak selection leads to a different exclusion list, which leads to a subsequent change in peak selection, and so on. The second effect thus magnifies the impact of small differences in peak selections early in the experiment.  Overall, the mismatch between reality and simulation underscores the importance of establishing a simulation baseline against which to compare the results of alternative peak selection procedures.

\subsection*{Variable-Duration Exclusion List}

\begin{figure*}
\centering
\begin{tabular}{cc}
\includegraphics[width=3.0in]{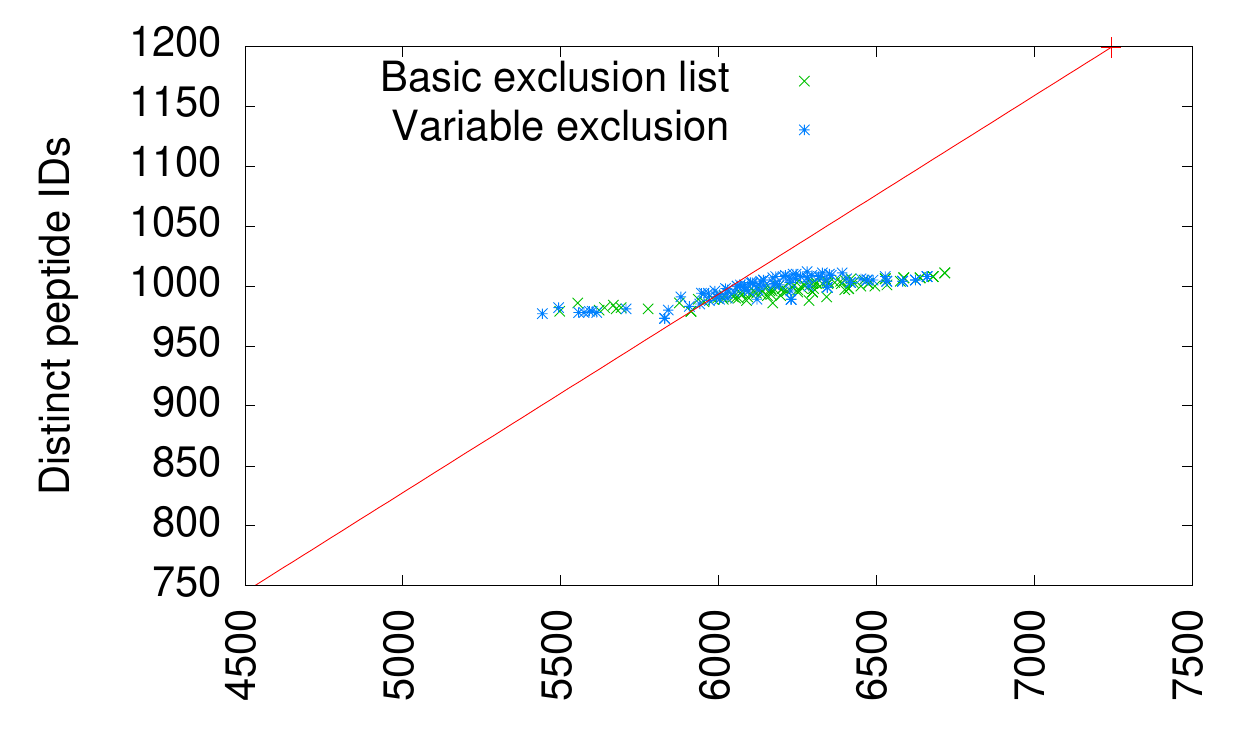} &
\includegraphics[width=3.0in]{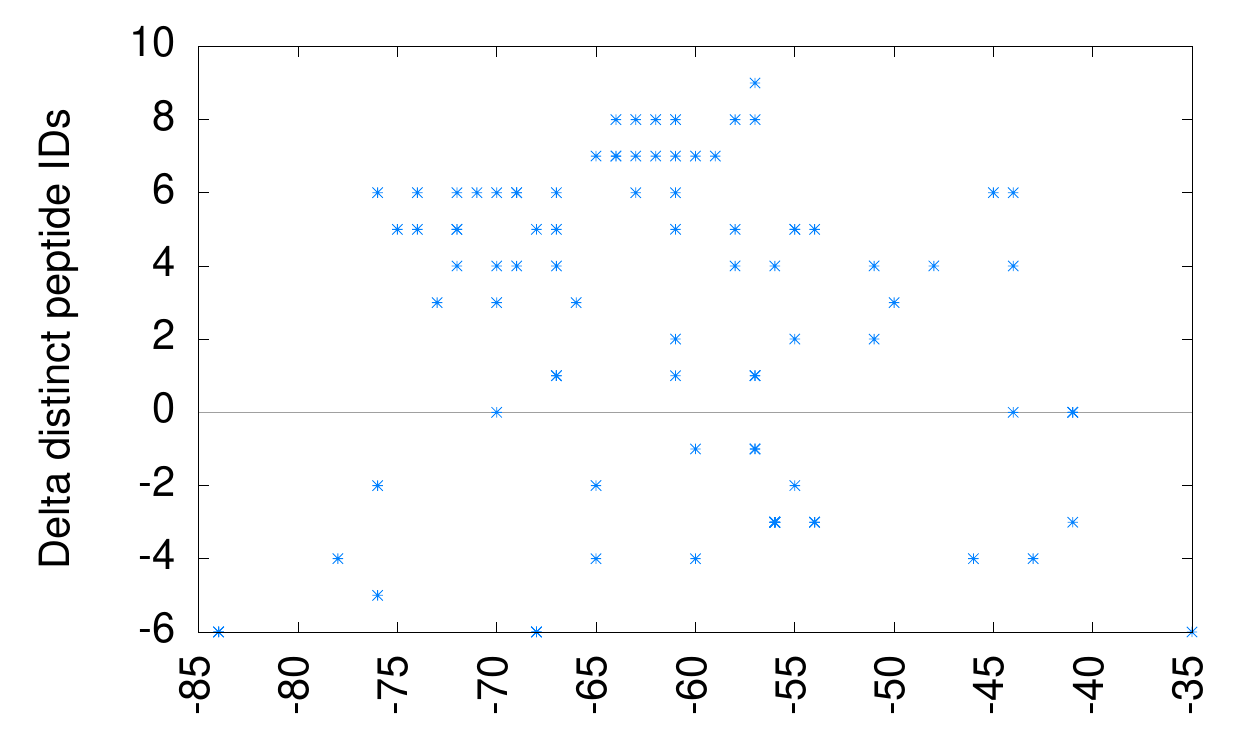} \\
\includegraphics[width=3.0in]{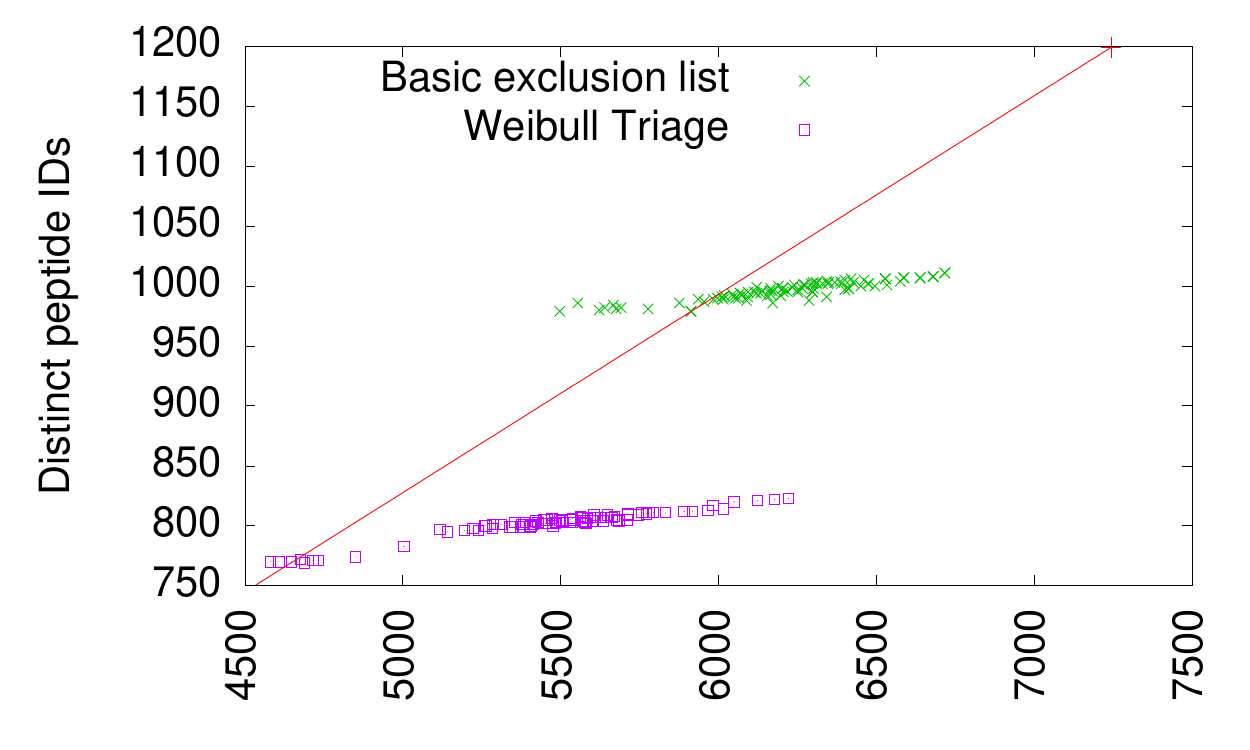} &
\includegraphics[width=3.0in]{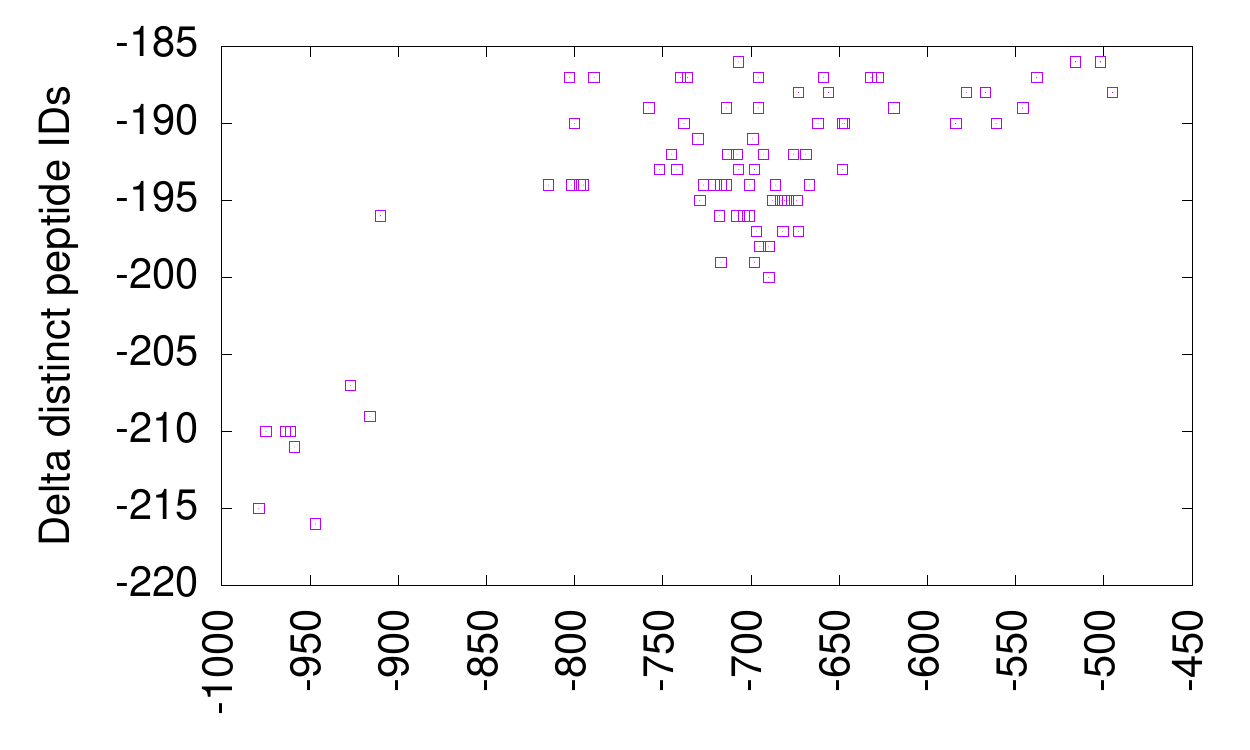} \\
\includegraphics[width=3.0in]{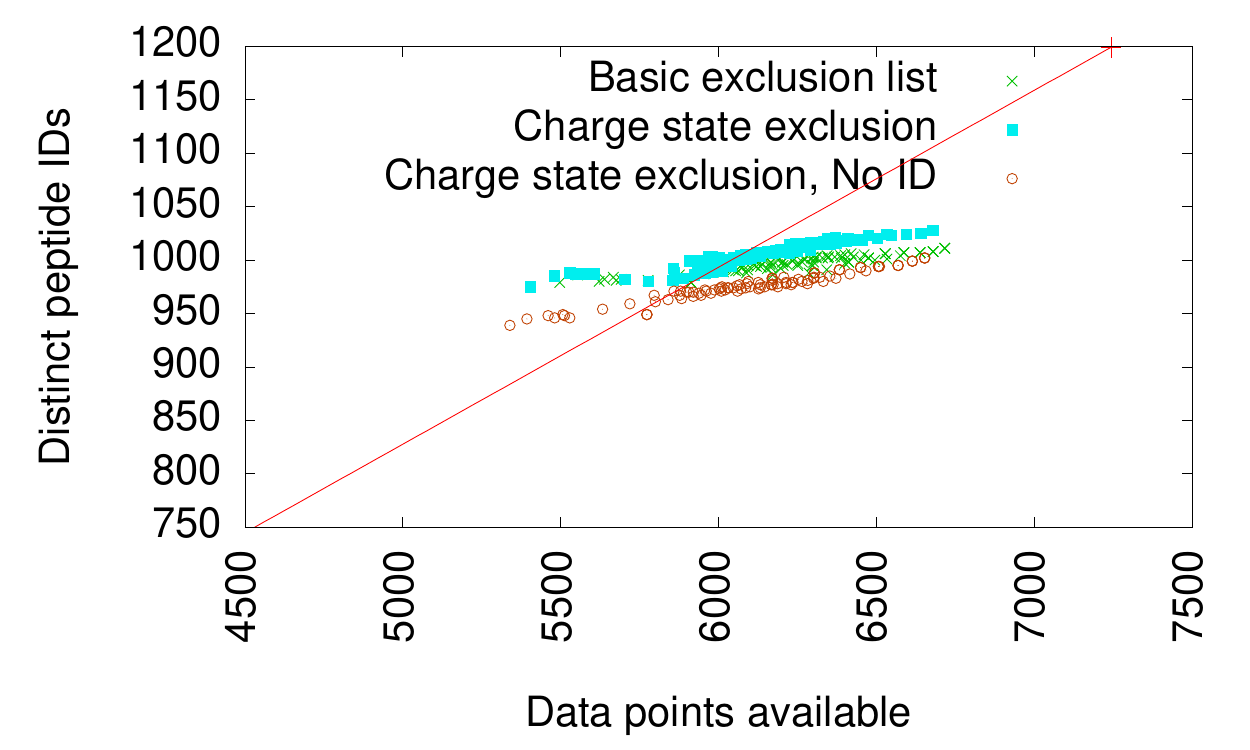} &
\includegraphics[width=3.0in]{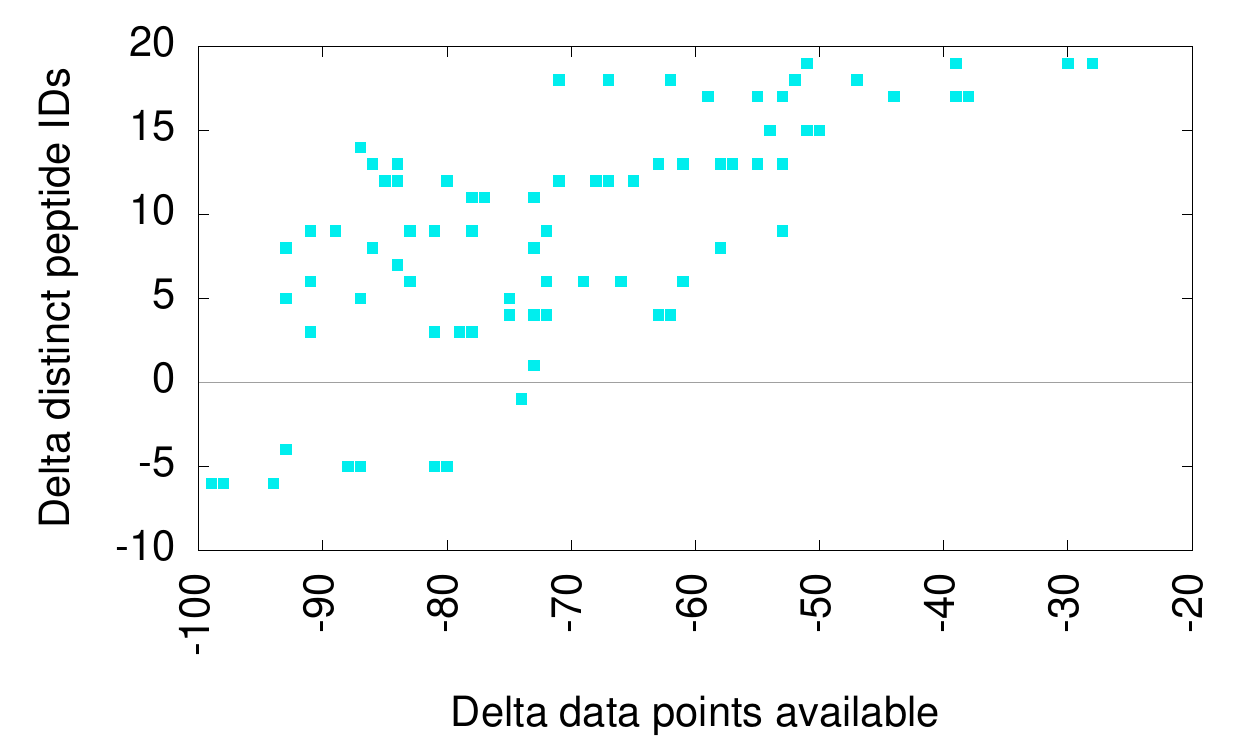} \\
\end{tabular}
\caption{{\bf Simulation results.}  In the figure, each left-hand panel shows the results of simulations, under
  varying tolerances, of a specified peak selection
  procedure. 
 Each plotted point represents a
  complete simulated mass spectrometry experiment under varying
  parameterizations of the simulator. Each panel plots the total number of
  distinct peptides identified as a function of the number of selected peaks for which the simulation successfully returned an MS2 spectrum. In each case, the red cross in the upper right represents the actual e10
  experiment, and the red line represents the ratio of identifications to total peaks
  acquired throughout the run of e10.
For comparison, each panel also includes the results for
  the basic exclusion list.  In the right-hand panels, for each
  setting of simulation tolerances, we plot the differences
  between a specified peak selection procedure and the basic exclusion
  list. The $y$-axis shows the difference in the number of distinct
  peptides identified, and the $x$-axis shows the difference in the
  number of selected peaks for which the simulator returned an MS2
  spectrum. 
  \label{figure:simulation}}
\end{figure*}

The hypothesis underlying our first peak-picking method is that the
basic exclusion list duration is not optimal for acquiring a diversity of
peptides, once the peptides are knowable. The 18-second exclusion list duration
represents a mean duration time, but performance could be improved if peaks corresponding to quickly-eluting peptides
came off the list quickly, whereas peaks corresponding to slowly-eluting peptides should remain on the list for a longer period of time. Based on this intuition, we reasoned that if we know the identity of the
selected peaks, then a repeated identification is a good indication that the peptide
is still eluting. In that case, a longer than normal exclusion list duration
should apply for that peak.

To implement this strategy, we gather peaks in order of intensity, just
as for the basic exclusion list, but we modify the decision to include or
exclude them. We allow a peak to be requeried after the normal
exclusion time. However, if a peak is requeried and is identified as the same
peptide as previously, the peak subsequently remains on the exclusion list for
three times longer.

The results from the variable duration exclusion list procedure are shown in the top two panels of Figure~\ref{figure:simulation}. We can see that for a
majority of simulation settings (most individual plotted points), the variable
exclusion list identifies slightly more distinct peptide than the simulated basic
exclusion list. However, the difference in performance is extremely modest: an
average increase of 1.9 distinct peptides out of an average total of 997 for
the basic exclusion list.

\subsection*{Weibull-score Triage}

The second peak-picking method that we investigated is based on the
hypothesis that if a peak is not confidently identified, then it may
be beneficial query the same peak a second time.  The method
subdivides spectra into three groups: (1) those with high scores that
do not need to be re-queried, (2) those with very low scores for which
a re-query is not likely to lead to a successful identification, and
(3) spectra with intermediate scores that should be re-queried.  This
hypothesis sets up a natural triage as follows. If a peak is
identified with high confidence or low confidence, then it should be
placed on the exclusion list according to the standard
protocol. Alternatively, if the peak is identified with middling
confidence, then we allow it to be requeried after a much shorter
interval.

We used a previously described confidence estimation metric, 
based on fitting a Weibull curve to the tail of the scoring distribution over
the candidate peptides \cite{klammer:statistical}. We employed this score on the theory that, if this method worked
well, we could introduce the Weibull-based confidence estimator into Tide while
keeping the performance high enough to meet the throughput and latency
requirements for real-time peptide identification.

We then applied the following specific protocol:
\begin{enumerate}
\item If a selected peak was identified with a $p$-value of $0.01$
according to the Weibull curve fit, then it was placed on the exclusion list
normally.
\item If a selected peak failed to be identified confidently ($p$-value $>0.1$),
  then it was also placed on the exclusion list normally.
\item If a selected peak was identified with intermediate confidence ($p$-value
  between $0.01$ and $0.1$), then it was placed on the exclusion list {\em only
    briefly} and was allowed to be requeried after 5 seconds.
\end{enumerate}

This approach performed noticeably poorly relative to the basic exclusion list---about 19\% fewer identifications than the basic exclusion list, averaged
across a variety of tolerances, albeit with 11\% fewer available simulation
peaks. Detailed results are shown in the middle two panels of Figure~\ref{figure:simulation}. We investigated other triage thresholds and obtained similarly poor results (data not shown).

\subsection*{Charge state exclusion}

The third peak-picking method we investigated is based on the fact that the same
peptide may elute roughly simultaneously at two different charge states. The
basic peak selection protocol may then end up choosing both such peaks. No
effort is made to prefer one over the other, because it may be easier to
identify the peptide at one charge state or the other, and the spectrometer does
not know {\it a priori} which is preferable. However, we hypothesize that if a peak is identified with high confidence at
one charge state, then we can reasonably exclude the corresponding peaks at the
other charge states as well, because they are likely to be the same peptide
species. If a peak is not identified with high confidence, then it may be
treated as before.

The results, shown in the bottom two panels of Figure~\ref{figure:simulation}, are somewhat more
promising than for the other peak picking methods we examined, but remain
modest: an average of 10 more unique peptide identifications (1\%) over the
simulated basic exclusion list. To be sure of the effect, we also show the
results (shown in red crosses) of simply excluding peaks at sister charge states
on a first-come-first-served basis, without regard to whether the identification
at the first attempted charge state was made confidently. These results are
worse than the basic exclusion list and show that real-time identification was
critical in gaining the additional identifications seen in the charge state
exclusion method.

\section*{Discussion}

Overall, our results point to one positive conclusion and one negative conclusion.  On the one hand, we have demonstrated that real-time assignment of peptides to fragmentation spectra is feasible, in principle.  
Much larger searches, such as non-enzymatic digest or searches
allowing multiple modifications per peptide, would require more memory
and more time than the above strategy provides. Hence, there certainly remain
limits to what can be achieved in a real-time context, but our analysis
brings closer together the notions of high-throughput and of real-time peptide identification.

On the other hand, our simulations have failed to provide evidence that real-time identifications yield significant value in the context of selecting MS1 peaks for fragmentation. In particular, we could not demonstrate that any of the three peak-selection strategies would yield more than a 1\% increase in unique peptide identifications relative to the standard exclusion list.

Among the peak-selection strategies that we tried, 
the Weibull-triage method fared poorly, while the other two
methods showed very slight improvements over the basic exclusion list, with the
charge-state exclusion being the most promising.
Thus, it seems that the
better-performing approaches rely on aggressively excluding peaks rather than
relaxing the exclusion criteria. Rather than requerying peaks that failed to be
identified previously, we are better off excluding those peaks that have any
likelihood of being a duplicate. Our interpretation is that this is because the
generally low identification rate of peptides is a stronger effect than that of
noise in the spectrum.

We were stymied in getting more positive peak selection results by at least three effects. The first was the insufficiency of data for a good simulation: many requested peaks did not have corresponding scans. We attempted to address this problem by aggregating over replicate experiments, but clearly, aggregating over a larger set of experiments would lead to a better simulation.  We also applied a two-dimensional success metric that attempts to correct for missing simulation data.

The second effect was that results were more sensitive to uninteresting parameters, such as simulation tolerances, than interesting ones.  To address this problem, we allowed the simulation parameters to vary, and we aggregated over all of those results.

The third effect was that the actual peak-picking strategy employed by the spectrometer is apparently more complex than a straightforward exclusion list. In principle, we could attempt to reverse engineer the mechanism of the on-board peak selector; however, the instability of the exclusion list argues against this approach.  Experimental peak selectors will necessarily deviate from the on-board peak selector's choices, so the subsequent effects of exclusion list instability are unavoidable in evaluating alternative peak selectors. It therefore makes sense in evaluation to use a baseline that is subject to the same instability.

\bibliographystyle{unsrt}
\bibliography{refs}
\end{document}